\documentclass[twocolumn,pre,aps,showpacs,groupedaddress]{revtex4-1}

\usepackage{amsmath,amssymb}
\usepackage{pstricks,placeins}
\usepackage{hyperref,url,epsfig}
\usepackage[latin1]{inputenc}
\psset{unit=1mm}

\def\<{\left<}
\def\>{\right>}
\def\ket|#1>{\left|#1\right>}
\def\bra<#1|{\left<#1\right|}
\def\elem<#1|#2|#3>{\left<#1\right|#2\left|#3\right>}
\def\({\left(}
\def\){\right)}

\begin{document}

\title{Topology and the Kardar-Parisi-Zhang universality class}

\author{Silvia N.\ Santalla$^{1,5}$}\email{silvia.santalla@uc3m.es}
\author{Javier Rodr\'{\i}guez-Laguna$^{2,5}$}
\author{Alessio Celi$^3$}
\author{Rodolfo Cuerno$^{4,5}$}
\affiliation{$^1$Departamento de F\'{\i}sica, Universidad Carlos III
  de Madrid, Spain\\
$^2$Departamento de F\'{\i}sica Fundamental, Universidad Nacional de
  Educaci\'on a Distancia (UNED), Spain \\
$^3$Institute of Photonic Sciences (ICFO), Barcelona, Spain \\
$^4$Departamento de Matem\'aticas, Universidad Carlos III de Madrid, Spain \\
$^5$Grupo Interdisciplinar de Sistemas Complejos (GISC)}

\date{April 25, 2016}

\begin{abstract}
We study the role of the topology of the background space on the
one-dimensional Kardar-Parisi-Zhang (KPZ) universality class.  To do
so, we study the growth of balls on disordered 2D manifolds with
random Riemannian metrics, generated by introducing random
perturbations to a base manifold. As base manifolds we consider cones
of different aperture angles $\theta$, including the limiting cases of
a cylinder ($\theta=0$, which corresponds to an interface with
periodic boundary conditions) and a plane ($\theta=\pi/2$, which
corresponds to an interface with circular geometry).  We obtain that
in the former case the radial fluctuations of the ball boundaries
follow the Tracy-Widom (TW) distribution of the largest eigenvalue of
random matrices in the Gaussian orthogonal ensemble (TW-GOE), while on
cones with any aperture angle $\theta\neq 0$ fluctuations correspond
to the TW-GUE distribution related with the Gaussian unitary ensemble.
We provide a topological argument to justify the relevance of TW-GUE
statistics for cones, and state a conjecture which relates the KPZ
universality subclass with the background topology.
\end{abstract}


\maketitle


\section{Introduction}

Growth is about geometry, even in the presence of noise. The
Kardar-Parisi-Zhang (KPZ) universality class, which describes the
fluctuations of growing one-dimensional interfaces,
\cite{Kardar_PRL86,Barabasi} is known to also describe the statistics
of the boundaries of balls with increasing radii on random
two-dimensional manifolds which are flat on average
\cite{Santalla_NJP15}. Remarkably, the KPZ class does not only entail
the values of the critical exponents, but also the full probability
distribution for the one-point and the two-point fluctuations, which
were initially conjectured and later shown to follow Airy processes
\cite{Praehofer_PRL00,Takeuchi_PRL10,Sasamoto_PRL10,Amir_CPAM11}, see
e.g.\ \cite{Corwin_RM12} for a recent review. Nonetheless, at this
level the class splits into two subclasses. In band geometry, i.e.,
for an interface with periodic boundaries, the local fluctuations are
ruled by the Tracy-Widom largest eigenvalue distribution associated
with the Gaussian orthogonal ensemble (TW-GOE) of random matrices
\cite{Calabrese_11}. On the other hand, if the interface has an
overall circular shape, the fluctuations are those characteristic of
the Gaussian unitary ensemble (TW-GUE). What is the origin of such a
splitting of the class into two topological flavors or subclasses?
Recent work on discrete growth models and the KPZ equation itself
\cite{Carrasco_NJP.14,Halpin-HealyJSP15} shows that, if the interface
is in a band geometry but the underlying substrate is growing, the
fluctuations are TW-GUE, just as in the circular case.  This shows
that the interface does not need to have a non-zero global curvature
for TW-GUE statistics to occur.

All these considerations point to relevant questions: what kind of
change takes place in the KPZ subclass when the topology of the base
manifold on which growth occurs is changed in a continuous way? What
are the relevant subclasses occurring? The possibility of exploring
the KPZ class on any Riemannian manifold was already put forward with
the proposal of a {\em covariant form} of the KPZ equation, which was
used to explore band and circular geometries simply by changing the
base manifold \cite{Laguna_JSTAT11,Santalla_PRE14}. It was shown that,
before reaching the KPZ behavior, the system explored a transient
state: a self-avoiding walk (SAW) or an Edwards-Wilkinson (EW)
crossover for band and circular geometry, respectively.  As a
particular case, in the absence of noise or diffusive terms one can
study the equation which merely propagates an interface with a
constant speed along the local normal direction ---related with the
level set equation in the case of the dynamics of function graphs
\cite{Sethian_book}---, which we call {\em Huygens equation}. If
applied to an infinitesimal circle, such an equation yields balls of
increasing radii around the central point.  In \cite{Santalla_NJP15},
such a Huygens equation was studied on random or disordered Riemannian
manifolds with short-range correlations, which are flat on average.
The dynamics of ball boundaries with increasing radii were shown to
fall into the KPZ universality class, the radial fluctuations
following the TW-GUE distribution. A relevant point is that
transients were absent in this case: KPZ universal behavior was
reached already for very short times.

\begin{figure*}
\epsfig{file=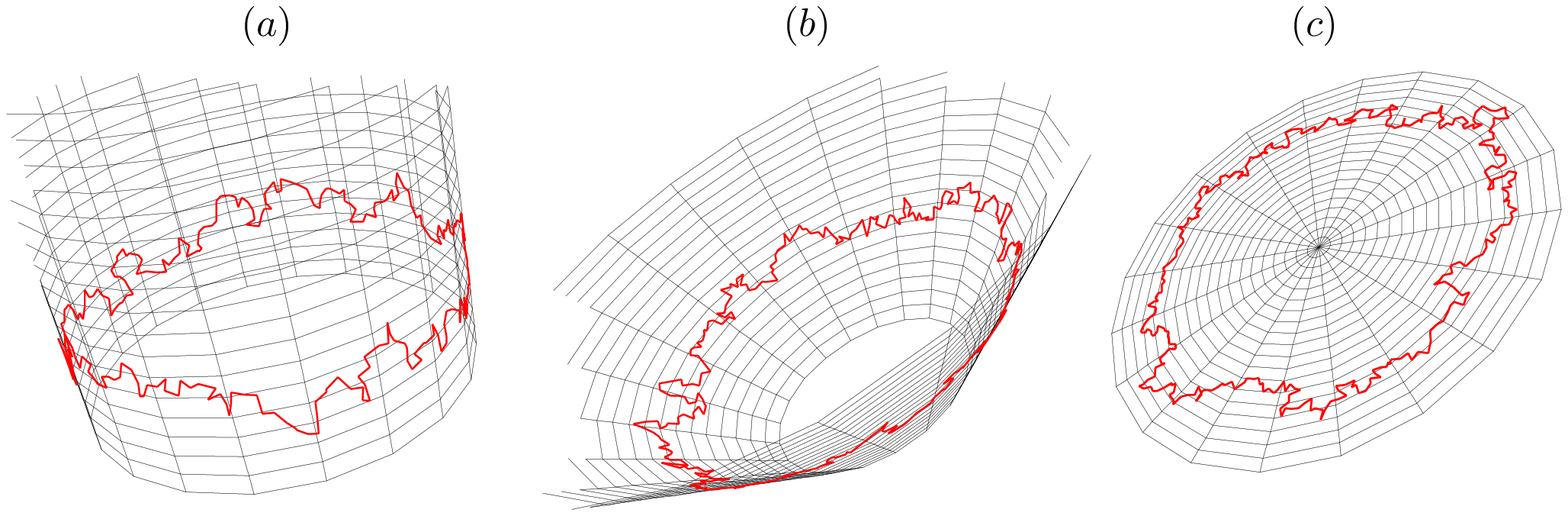,width=18cm}
\caption{Ball boundaries on a random manifold whose background metric
  is (a) a cylinder, (b) a cone, and (c) a plane. These interfaces
  have been generated using the numerical algorithm described in
  Sec.\ \ref{sec:numerics}.}
  \label{fig:illust_3d}
\end{figure*}

In this work we study the effect of topology on the subclass structure
of the KPZ universality class, by exploring the interface fluctuations
for growing balls on different types of random Riemannian
manifolds. More concretely, we study the interfaces developed by the
Huygens equation on cones of different opening angles, including the
limiting cases of the cylinder and the plane, which is the case
studied in \cite{Santalla_NJP15}.  See Fig.\ \ref{fig:illust_3d} for
an illustration.

Our overall conclusion is that TW-GOE statistics are characteristic of
the cylinder, TW-GUE behavior occurring for cones of arbitrary
aperture angles $\theta$, including the plane ($\theta=\pi/2$). Hence,
a change takes place in the KPZ universality subclass between TW-GUE
and TW-GOE as the aperture angle of the base cone manifold is changed,
for $\theta=0$.  Transitions among the various KPZ subclasses have
been previously explored, although mostly when considering particular
initial conditions which are such that, at long times, the interface
divides into spatial regions in which statistics are of one or the
other subclass.  See e.g.\ \cite{Borodin2008} for the case of the
totally asymmetric simple exclusion process (TASEP) model with an
initial condition where particles are placed at the even integers. Or
the KPZ equation with a double-wedge initial condition or,
equivalently, a directed polymer on a half-space with an end-point
fixed \cite{LeDoussal2014}. In our present case, the statistics are
homogeneous throughout the system and change abruptly from TW-GOE to
TW-GUE as soon as the aperture angle is non-zero. Such a result
complements those obtained in growing systems with a band geometry
\cite{Carrasco_NJP.14,Halpin-HealyJSP15}, in the sense that these two
are the only relevant subclasses in the presence of this type of
topological changes.

This paper is structured as follows. Section \ref{sec:theory}
discusses our general framework: the covariant KPZ equation and
Huygens equation, considered on random conformal deformations of a
given base manifold. In section \ref{sec:geometry} we describe the
parametrization that we will use for the cylinder, cones and plane, an
the base metric. In section \ref{sec:numerics} we discuss our
numerical simulations of interfaces on random cones, the critical
exponents, and the radial fluctuations. The fact that all cones have
TW-GUE radial fluctuations is justified in section
\ref{sec:analysis}. Our conclusions and ideas for further work are
finally outlined in section \ref{sec:conclusions}.


\section{From the covariant KPZ equation to random metrics}
\label{sec:theory}

In previous works \cite{Laguna_JSTAT11,Santalla_PRE14}, we have
proposed an extension of the KPZ equation for which all terms are
defined in a covariant manner, i.e., the equation has the same form
when expressed on any background metric. The equation expresses the
evolution of a closed simple curve representing an interface. Each
point ${\vec r}$ on the curve moves along the local normal direction,
with a velocity affected by three different terms:

\begin{equation}
\partial_t {\vec r}=\left[ A_0 + A_1 k({\vec r}) + A_n \eta({\vec r})\right]
\, \vec n(\vec r).
\label{eq:ckpz}
\end{equation}

\noindent Here, $\vec n$ is the local unit normal vector, $k$ is the
geodesic curvature, and $\eta$ is a zero-average Gaussian noise,
uncorrelated both in time and along the interface. The constants
$A_0$, $A_1$, and $A_n$ are free parameters, which characterize,
respectively, irreversible growth, surface tension, and fluctuations
in the growth events. In fact, this interface can develop
self-intersections. Thus, Eq.\ \eqref{eq:ckpz} must be supplemented
with an algorithm to treat them. A convenient choice is to remove
always the smaller component \cite{Laguna_JSTAT11,Santalla_PRE14}.

In \cite{Santalla_NJP15} we focused on the simplest case of
Eq.\ \eqref{eq:ckpz} with $A_1=A_n=0$, which we call the {\em Huygens
  equation}, namely,

\begin{equation}
\partial_t \vec r= \vec n(\vec r),
\label{eq:huygens}
\end{equation}
because it simply {\em propagates} any closed curve outwards, in a way
which is similar to Huygens' principle for the propagation of a
wavefront \cite{Sethian_book}. If our initial curve is an
infinitesimal circumference around point $X_0$, then the evolution of
our interface will be given by a set of {\em balls} on this metric,
with linearly increasing radii. In \cite{Santalla_NJP15} we applied
Eq.\ \eqref{eq:huygens} to the study of the growth of balls on
two-dimensional {\em random} manifolds with smooth enough random
metrics, which are flat on average and have short-range correlations.

In the present work we lift the condition that the random metrics need
to be flat on average. Let us consider any background metric, given by
the metric tensor field $g_0(x,y)$. We can introduce an ensemble of
metrics through

\begin{equation}
g(x,y)=\nu(x,y) g_0(x,y) ,
\label{eq:random_metric}
\end{equation}
where $\nu(x,y)$ is a smooth enough random field with uniform average
and short-range correlations (as measured by the $g_0$ metric). This
means that the metric $g_0(x,y)$ is subject to a random conformal
transformation or, alternatively, that we consider an {\em optical
  metric} on the base manifold, with a position-dependent index of
refraction.


\section{Cylinder, cones, and plane}
\label{sec:geometry}

Let us address the study of the statistical properties of interfaces
generated by the Huygens equation \eqref{eq:huygens} on random
conformal deformations of a given base Riemannian manifold $g_0(x,y)$,
as expressed by Eq.\ \eqref{eq:random_metric}. The division of the KPZ
class between band geometry and circular geometry can be recast in our
Riemannian geometry language by stating that band geometry refers to
propagation of Huygens equation on a cylinder, while circular geometry
refers to propagation on a plane. Thus, for a random metric based on
the plane, the results of \cite{Santalla_NJP15} show that, as
expected, the radial fluctuations obey TW-GUE statistics. On a random
metric based on the cylinder, if we set up as initial condition a
curve which wraps around it, the ensuing interface fluctuations should
follow the TW-GOE distribution.

\begin{figure}
\epsfig{file=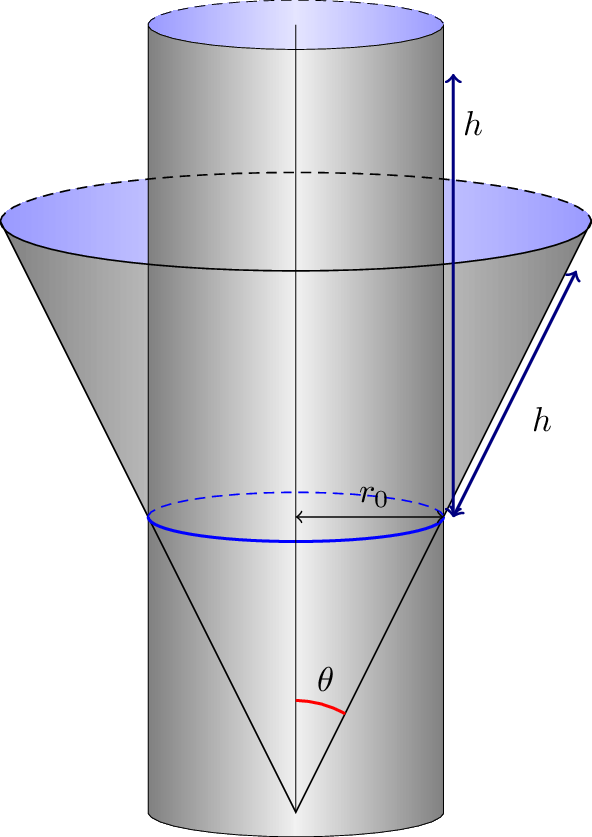,width=6cm}
\caption{Illustration for our family of conical surfaces, parametrized
  by $\theta$, the angle between the cone axis and generatrix. They
  are all forced to coincide on a base circumference of radius $r_0$,
  marked with the blue line. Quasi-polar coordinates are defined by
  using a radius $r=r_0+h$.}
  \label{fig:illust_cone}
\end{figure}

Let us define a natural family of surfaces which interpolates between
the cylinder and the plane: a set of cones of increasing opening angle
$\theta$ between the axis and the axis and the generatrix, with
$\theta=0$ for the cylinder and $\theta=\pi/2$ for the plane. See
Fig.\ \ref{fig:illust_cone} for an illustration. The cone can be
understood as a plane from which a wedge of angle $2\pi(1-\sin\theta)$
has been removed. We will address the following question: how does the
distribution for the normal fluctuations of the interface interpolate
between TW-GOE for the random metric on the cylinder and TW-GUE for
the random metric on the plane?

Cones are surfaces with zero Gaussian curvature $K$ everywhere except
at the vertex. The integral of $K$ over any domain containing the
vertex is always the same, and equal to the {\em angular defect}
$\Delta=2\pi(1-\sin \theta)$ \cite{Singer_Thorpe,Devadoss}. The sum of
the angles of any geodesic triangle containing the vertex will be
$\pi+\Delta$. In fact, there is a stronger version of this statement,
that is a consequence of the Gauss-Bonnet theorem:

\begin{equation}
\int_\gamma k_g\; ds = 2\pi\sin \theta .
\label{integrated_curvature}
\end{equation}
Here, $k_g$ is the geodesic curvature of any curve $\gamma$
surrounding the vertex. In the case of a random metric based on the
cone, Eq.\ \eqref{integrated_curvature} will be modified by
fluctuations. Yet, it shows that the integral of the geodesic
curvature is a conserved quantity on average, and we can expect some
observables of our interfaces to depend on $\theta$.

\subsection{Coordinates and metric on the cones}
\def\tx{{\tilde x}}
\def\ty{{\tilde y}}
\def\tr{{\tilde r}}

Let us describe our cone manifolds in detail, starting with their
embedding in 3D and moving to an intrinsic
chart. Fig.\ \ref{fig:illust_cone} shows the surfaces embedded in 3D
space, the $(X,Y,Z)$ coordinates of an arbitrary point on one of these
surfaces being given by

\begin{subequations}
\begin{eqnarray}
X &=& \left( r_0+h\sin\theta \right) \cos\phi, \\
Y &=& \left( r_0+h\sin\theta \right) \sin\phi, \\
Z &=& h \cos\theta,
\end{eqnarray}
\label{eq:conos_reales}
\end{subequations}
where we have made the cones coincide on a base circumference of
radius $r_0$ (the thick blue line in Fig.\ \ref{fig:illust_cone}) for
all $\theta$, $h$ is the distance of the point to the base
circumference, and $\phi$ is the azimuthal angle.  Let us choose a
{\em quasi-polar coordinate chart} on the cones, in which each point
is given by the pair $(r,\phi)$, with $r=r_0+h$. Thus, the base
circumference will be described as $r=r_0$ on all the cones. We can
now consider the metric for the cones expressed on these coordinates,

\begin{equation}
ds^2 = dr^2 + \rho^2(r)d\phi^2,
\label{eq:metrica_polar}
\end{equation}
where $\rho(r)=r_0+h\sin\theta=r_0+(r-r_0)\sin\theta$ is the distance
to the axis of the cone. The limit case of the cylinder ($\theta=0$)
yields

\begin{equation}
ds^2=dr^2+ r_0^2d\phi^2.
\label{eq:metrica_polar_cyl}
\end{equation}
Similarly, for the plane ($\theta=\pi/2$) we have

\begin{equation}
ds^2=dr^2+r^2d\phi^2.
\label{eq:metrica_polar_plane}
\end{equation}

Despite the simplicity of this quasi-polar metric, we prefer to
introduce a new Cartesian-like chart. The reason is to avoid the need
for periodic boundary conditions in the azimuthal angle. Let us define
$x$ and $y$ as

\begin{subequations}
\begin{eqnarray}
x &=& r\cos\phi, \\
y &=& r\sin\phi.
\end{eqnarray}
\label{eq:coor_planas}
\end{subequations}
Geometrically, the Cartesian-like coordinates $(x,y)$ express a
mapping of the cone on the plane containing the base circumference, in
which distances to this base curve are preserved. In these {\em
  quasi-Cartesian coordinates}, the metric can be written as

\begin{widetext}
\begin{subequations}
\begin{eqnarray}
 g_{xx} &=&
{ 2r_0 y^2 \left(r-r_0\right) \left(\sin\theta-\sin^2\theta\right) +
 r^4 \sin^2 \theta + \left( r_0^2 y^2+r^2 x^2 \right) \cos^2\theta
 \over r^4 }, \\
 g_{yy} &=&
{ 2r_0 x^2 \left(r-r_0\right) \left(\sin\theta-\sin^2\theta\right) +
 r^4 \sin^2 \theta + \left( r_0^2x^2+r^2y^2 \right) \cos^2\theta
 \over r^4}, \\
 g_{xy} &=& g_{yx} =
{ xy [ \left(r^2-r_0^2\right) \cos^2\theta - 2r_0 \left(r-r_0\right)
\left(\sin\theta-\sin^2\theta\right)] \over r^4}.
\end{eqnarray}
\label{eq:metrica}
\end{subequations}
\end{widetext}

Our numerical simulations will be performed on the $(x,y)$ plane, using the base
metric described by Eq.\ \eqref{eq:metrica}.


\section{Numerical Simulations and Results}
\label{sec:numerics}

In this section we describe our numerical simulations of the evolution
of the base circumference $x^2+y^2=r_0^2$ under Huygens equation
\eqref{eq:huygens}, supplemented with the rule of self-intersection
removal, on a random metric of the form \eqref{eq:random_metric},
i.e., a random conformal perturbation of the metric $g_0$. In turn,
$g_0$ will be one of our cone metrics, given by
Eq.\ \eqref{eq:metrica_polar} in (quasi-)polar coordinates or by
Eq.\ \eqref{eq:metrica} in (quasi-)Cartesian coordinates.

We have extended the algorithm described in
Ref.\ \cite{Santalla_NJP15} in order to work on random conformal
deformations of any given base Riemannian manifold. Let us summarize
the algorithm. The interface is considered to be a piecewise linear
simple curve, with an adaptive number of points: if two points
separate beyond a certain threshold $\ell_{max}$ (in the base metric
$g_0$), a new point is included mid-way \cite{Laguna_JSTAT11}. In all
cases, we take $\ell_{max}=0.05$. Each segment of the interface
determines a tangent vector $\vec t$ along the interface curve. We
make it evolve along the local normal direction $\vec n$. In order to
determine $\vec n$, we require the local metric tensor, $g(\vec
r)$. This is obtained, via Eq.\ \eqref{eq:random_metric}, by
multiplying the local metric tensor of the base manifold by a random
conformal factor, $\nu(\vec r)$. Then, we solve the equation $\vec t
\perp_g \vec n$, i.e., $g_{\mu\nu}(\vec r)t^\mu n^\nu=0$. The
propagation of each segment at each time-step ($\Delta t=0.005$) is
performed in a straightforward way, but the evolution equation is
supplemented with an algorithm in order to detect self-intersections
\cite{Laguna_JSTAT11}. As mentioned above, the smaller component is
always removed so that the interface remains a simple curve at all
times.

\begin{figure}
\epsfig{file=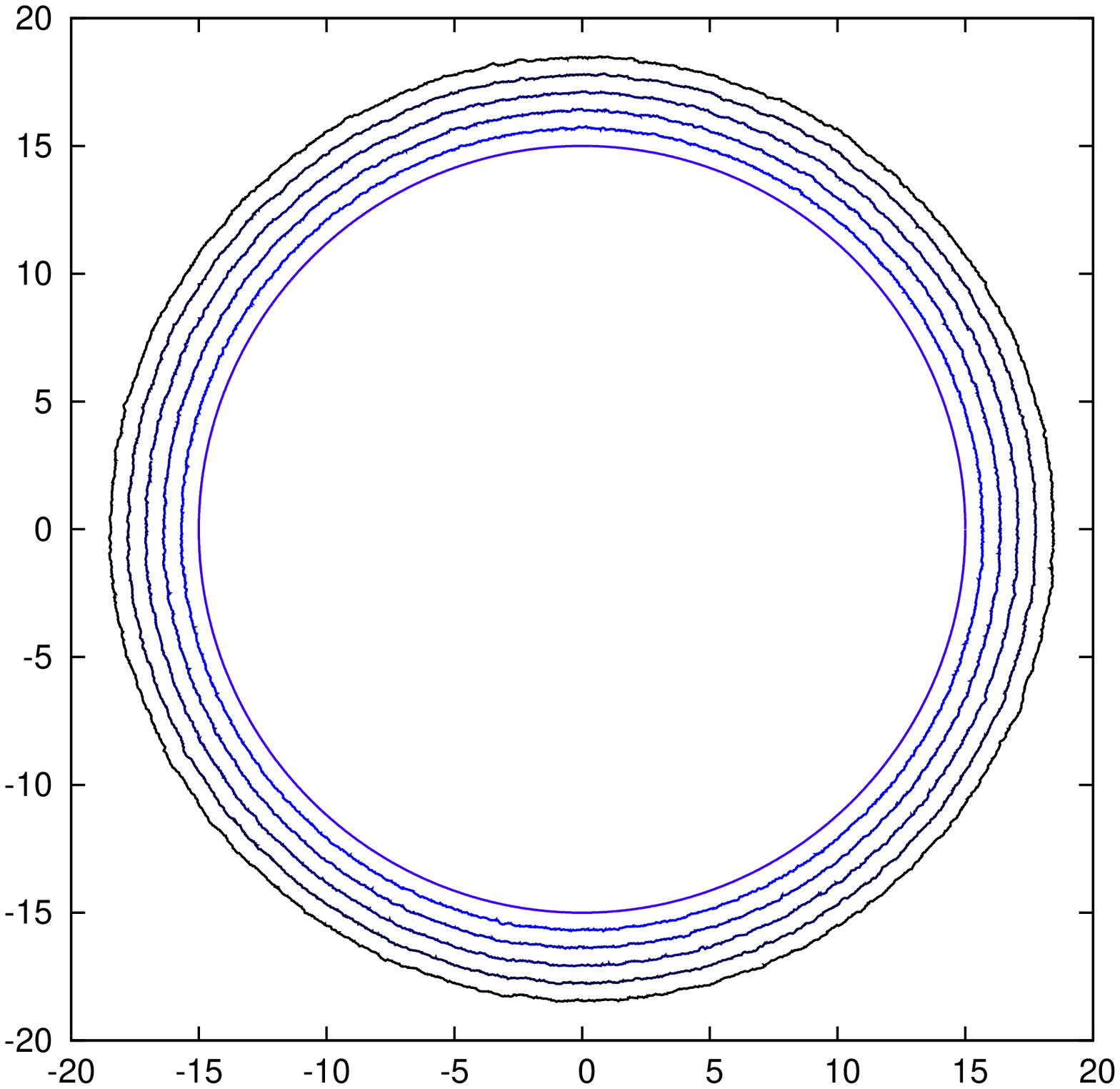,width=4cm,angle=270}
\epsfig{file=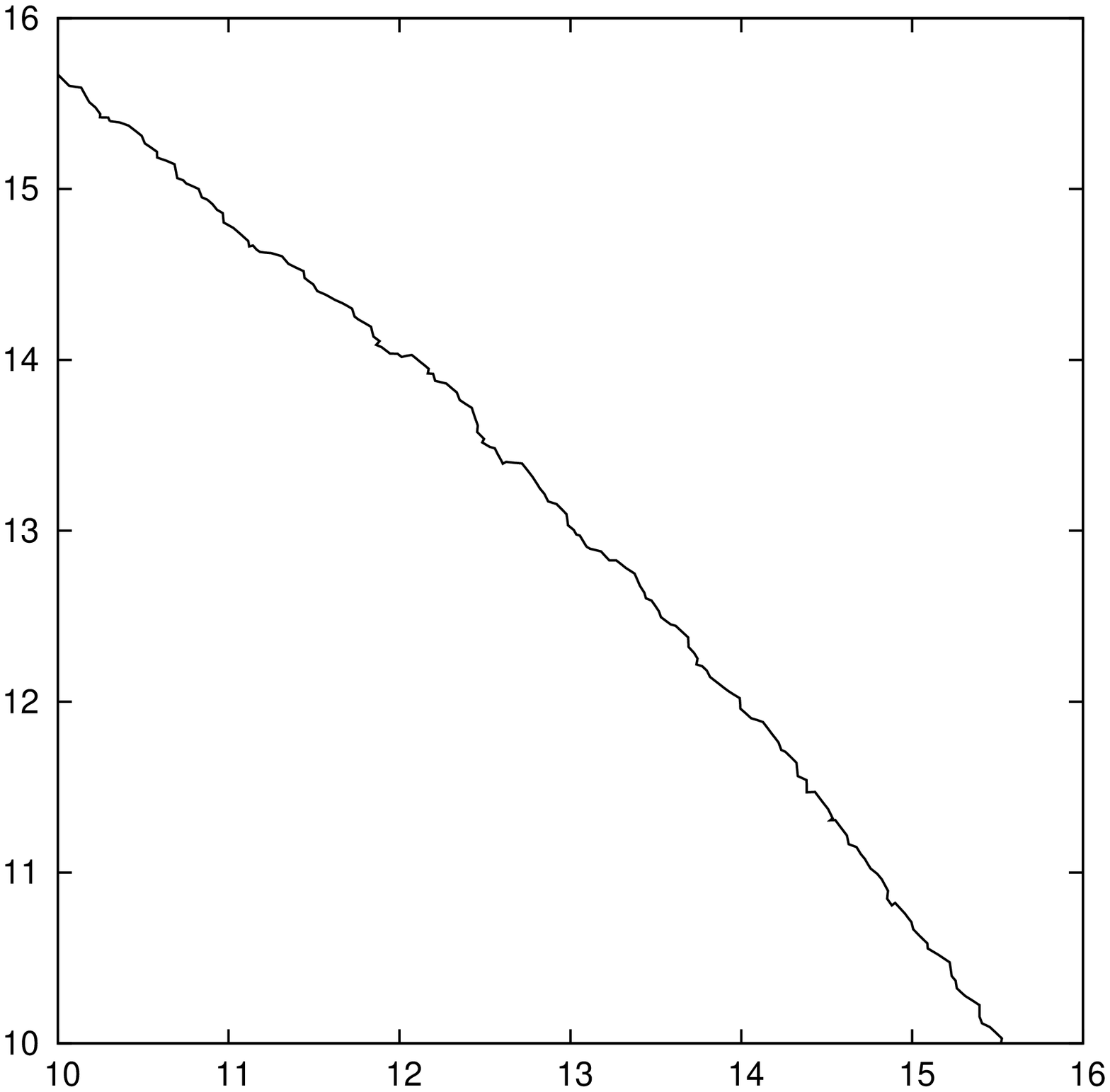,width=4cm,angle=270}
\epsfig{file=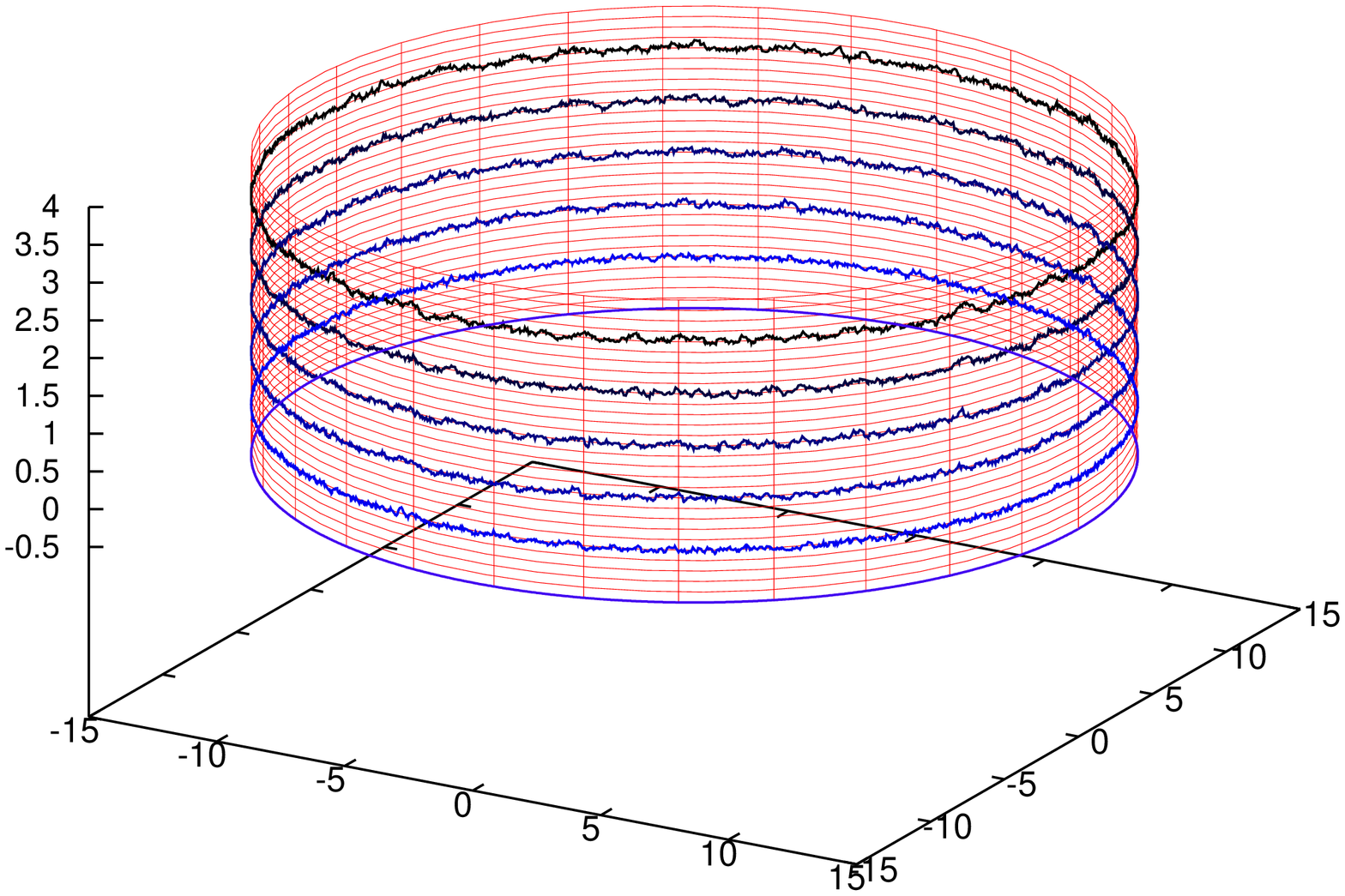,width=6cm,angle=270}
\epsfig{file=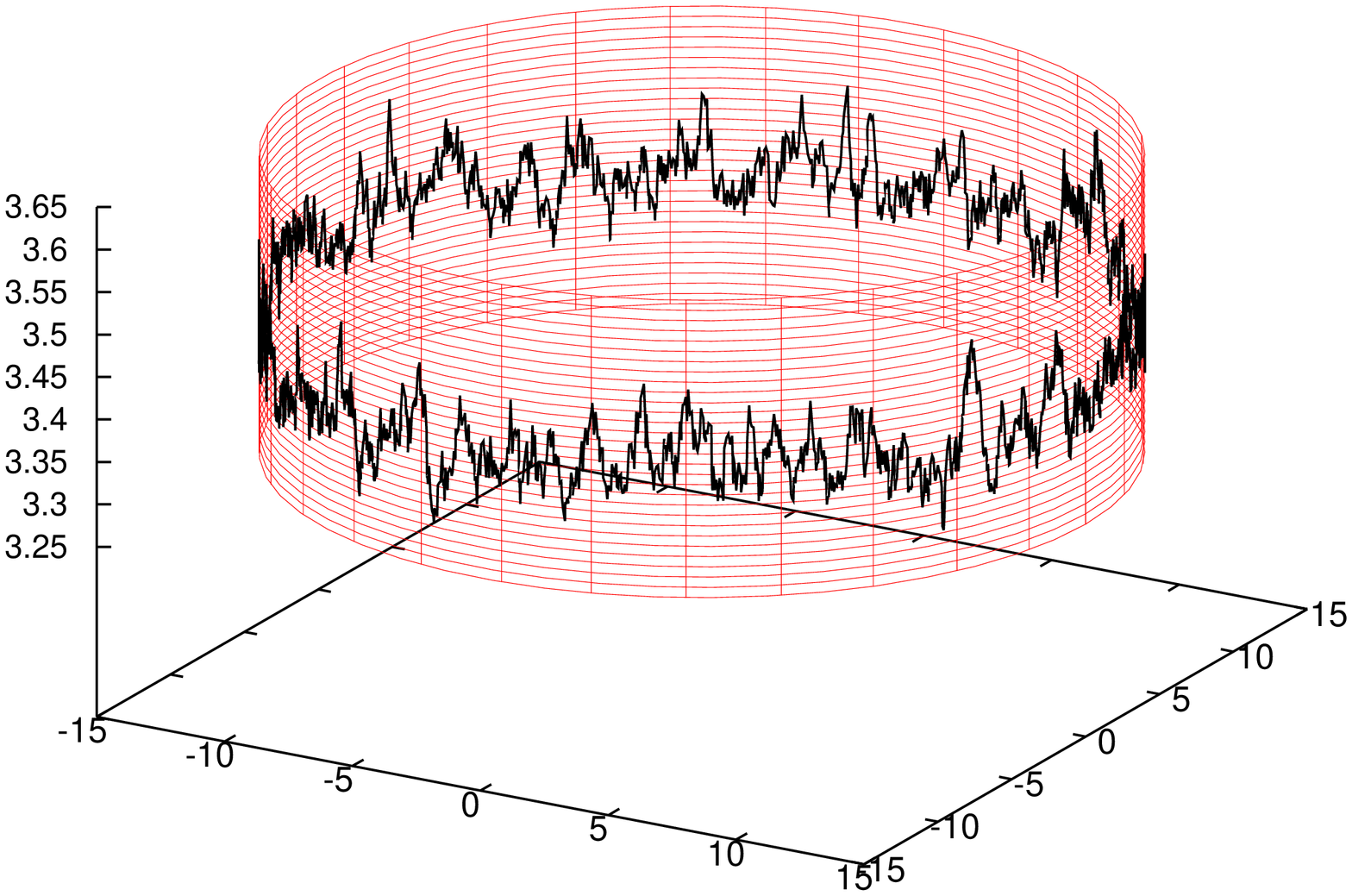,width=6cm,angle=270}
\caption{Interfaces on a cylinder with $r_0=15$. Top panels: profiles
  in $(x,y)$ coordinates. The top-right panel shows a zoom of the
  outermost profile in the top-left panel. Medium panel: profiles on
  the 3D cylinder. The simulation times are $t=0$, $4$, $8$, $12$,
  $16$ and $20$, bottom to top.  Bottom panel: enlargement of the
  $t=20$ profile shown in the center panel.}
\label{fig:cyl}
\end{figure}

\begin{figure}
\epsfig{file=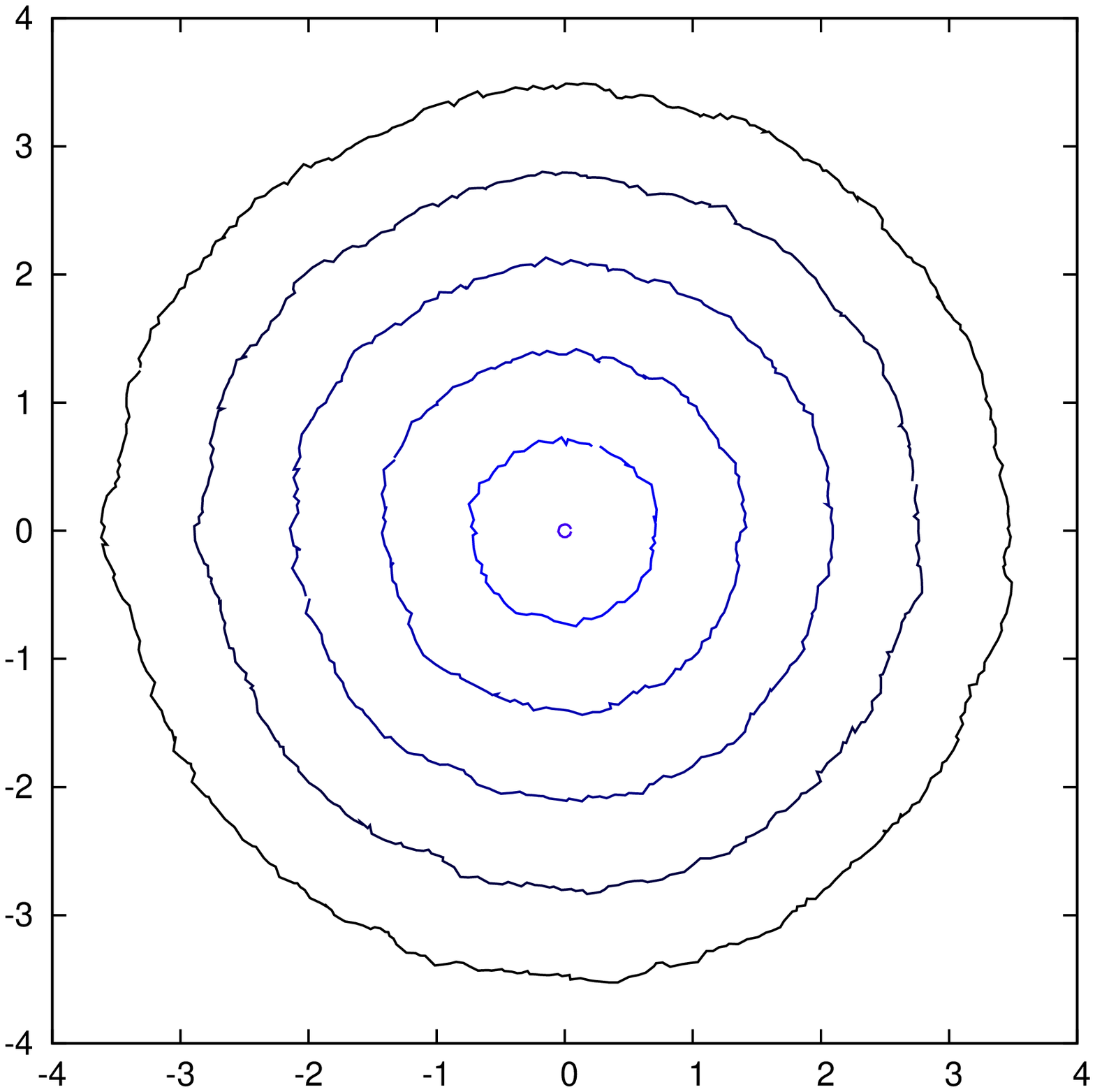,width=4cm,angle=270}
\epsfig{file=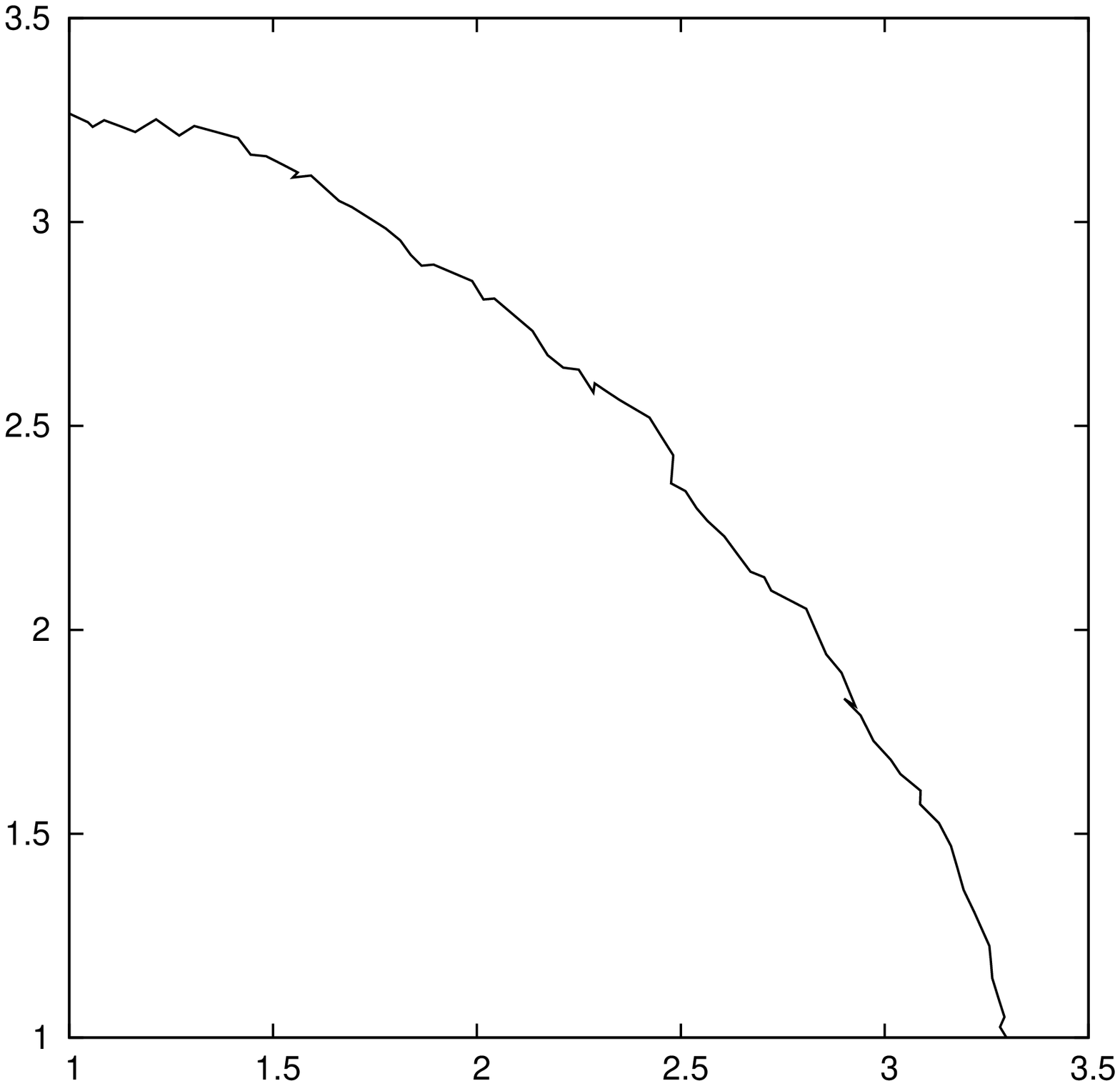,width=4cm,angle=270}
\epsfig{file=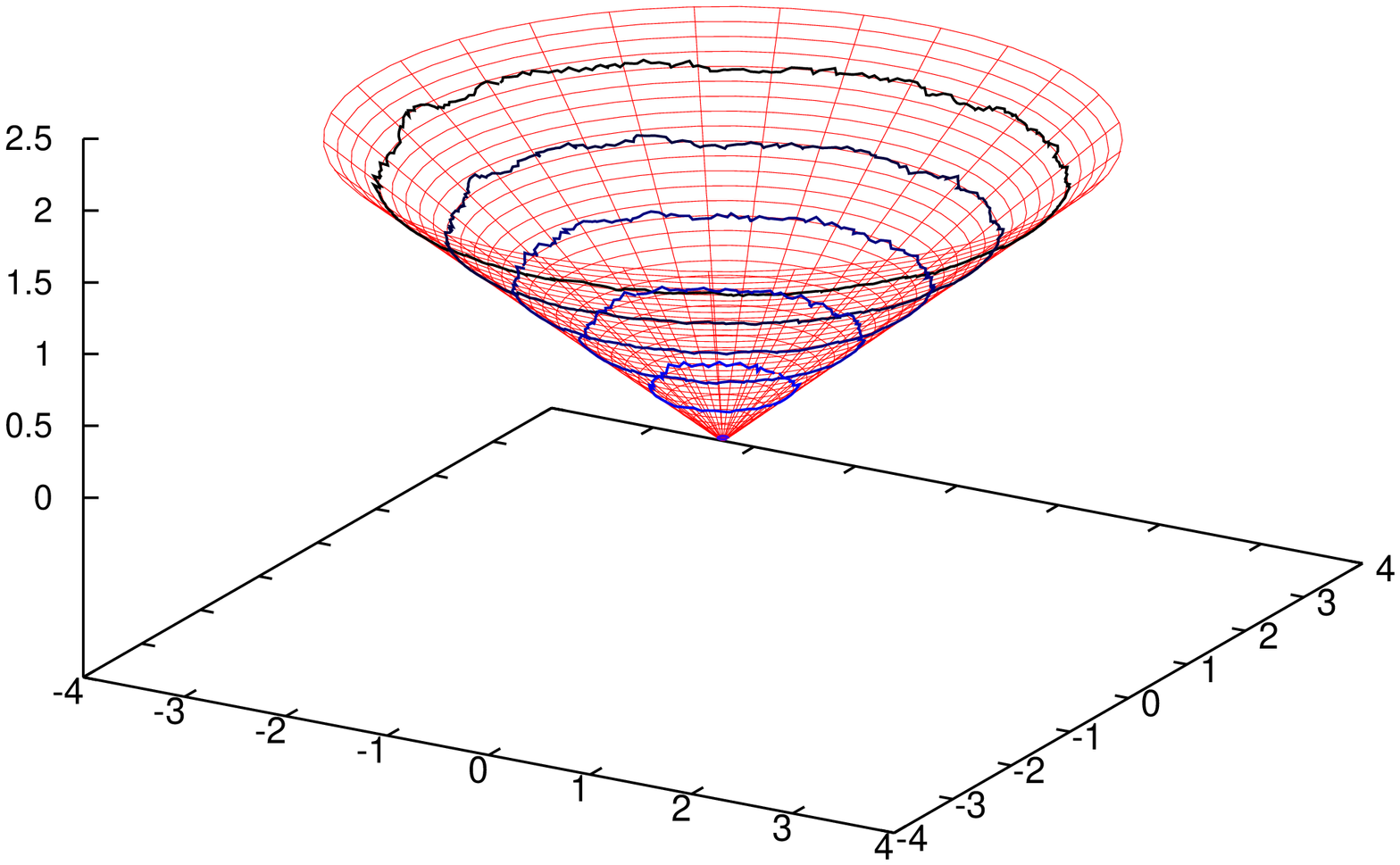,width=6cm,angle=270}
\epsfig{file=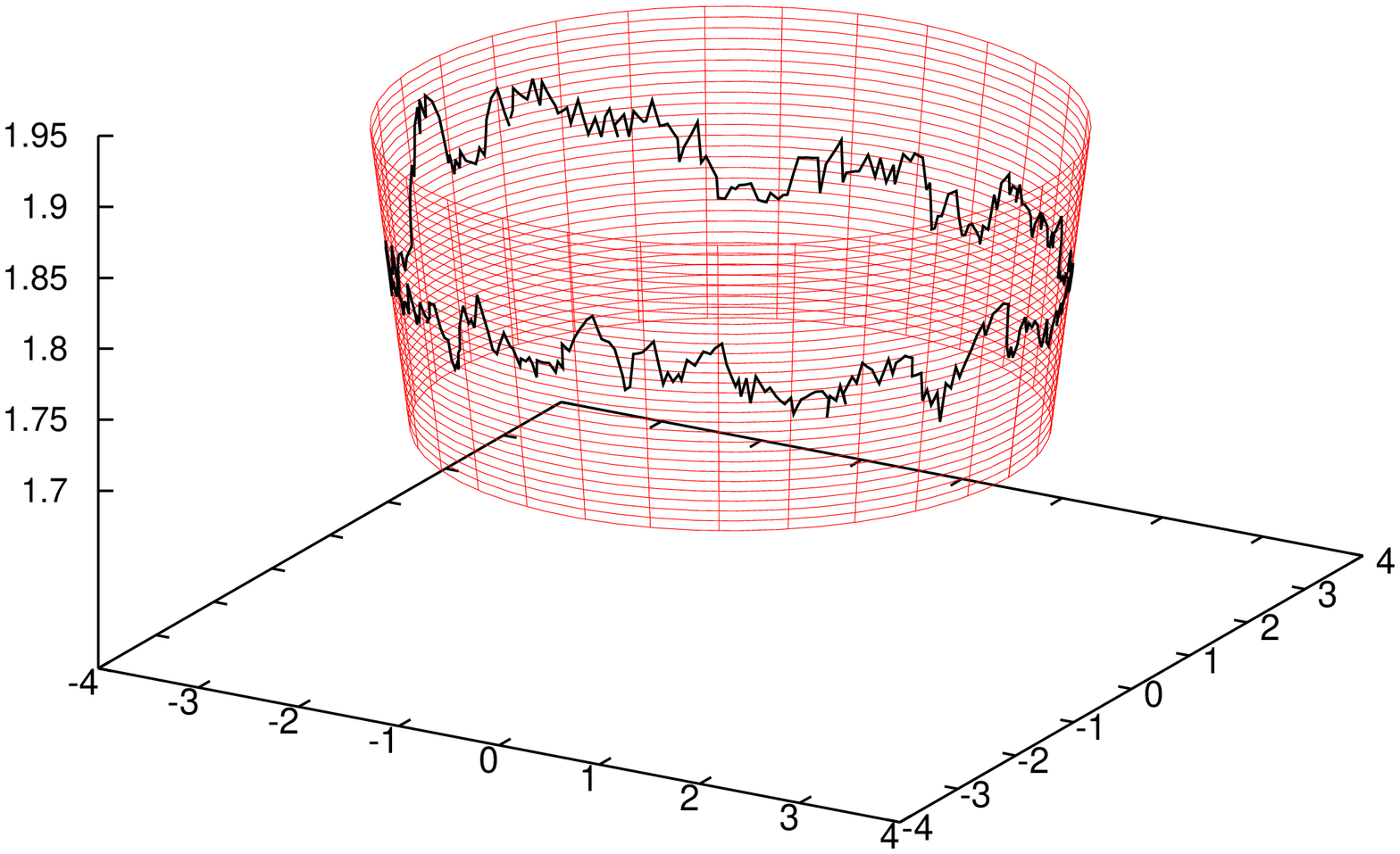,width=6cm,angle=270}
\caption{Interfaces on a cone with $\theta=\pi/4$ and $r_0=0.01$. Top
  panels: profiles in $(x,y)$ coordinates. The top-right panel shows a
  zoom of the outermost profile in the top-left panel. Medium panel:
  profiles on the 3D cylinder. The simulation times are $t=0$, $4$,
  $8$, $12$, $16$ and $20$, bottom to top.  Bottom panel: enlargement
  of the $t=20$ profile shown in the center panel.}
\label{fig:cone}
\end{figure}

Figures \ref{fig:cyl} and \ref{fig:cone} show some profiles obtained
by our simulations, for a cylinder and for a cone with $\theta=\pi/4$,
respectively. The initial radius is $r_0=15$ for the cylinder and
$r_0=0.01$ for the cone. The random conformal factor $\nu(\vec r)$ is
chosen as an uniform random deviate in $[1/20,1]$. In both figures,
the top panel shows the ball profiles as obtained in the $(x, y)$
coordinate chart. The top-right panel is a zoom of a single
profile. It can be noticed that the cylinder has much smaller
fluctuations, as we will explain shortly. The center panels show how
the previous interfaces fit on the original manifolds, the cylinder
and the cone. The bottom panel, in both cases, shows the interface
evolved up to the same time, $t=20$.

The bottom panels of Figs.\ \ref{fig:cyl} and \ref{fig:cone} show
that, in their 3D representation, the interfaces have comparable
roughness. The apparent suppression of fluctuations in the $(x,y)$
representation for the cylinder is due to the form of its metric,
Eq.\ \eqref{eq:metrica_polar_cyl}. Notice that it has a quasi-polar
form, in which distances along the azimuthal line do not grow with the
radius. Thus, a given fixed interface which is parallel-transported
upwards along the cylinder will appear less and less rough.

\subsection{Critical exponents}

As described in Ref.\ \cite{Santalla_NJP15}, ball boundaries on a
flat-average random metric of the form of
Eq.\ \eqref{eq:random_metric} follow the Family-Vicsek Ansatz when
considered as interfaces. Specifically, the roughness of the ball
boundary, as measured in the Euclidean metric, grows with time as a
power-law, $W(t) \sim t^\beta$, and so does the correlation length
along the interface, $\xi(t)\sim t^{1/z}$. Moreover, in the case
studied in \cite{Santalla_NJP15}, the values of the critical exponents
were shown to be those of the Kardar-Parisi-Zhang universality class,
$\beta=1/3$ and $1/z=2/3$.

\begin{figure}
\epsfig{file=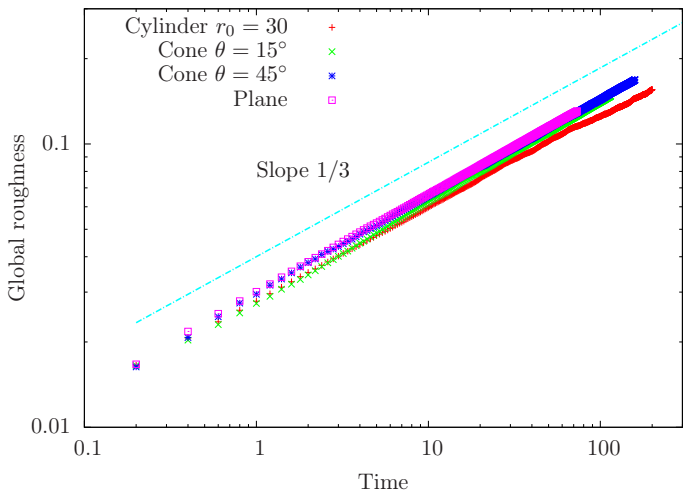,width=9cm}
\caption{Average roughness $W$ as a function of time for interfaces
  grown with Huygens equation \eqref{eq:huygens} on conformal random
  deformations of metrics corresponding to a cylinder of radius
  $r_0=30$, cones with opening angles $\theta=15^\circ$ and
  $45^\circ$, and a plane. In all cases, the roughness exponent is
  close to $1/3$.}
\label{fig:roughness}
\end{figure}

Let us now consider the interfaces produced by Huygens equation
\eqref{eq:huygens} on our random cones. The average shape of the ball
boundary for any given time is expected to be a circumference of
radius proportional to $t$. Although we do not have a proper shape
theorem for our general case, see \cite{Kesten.08,LaGatta.10} for some
rigorous shape theorems in particular manifolds. We define the
roughness of a curve $W$ as the expected magnitude of the normal
deviations of the actual interface from its best-fit circumference
centered at the origin. Notice that distances along the radial
direction in the $(x,y)$ chart can be computed in an Euclidean
setting. Fig.\ \ref{fig:roughness} shows this measurements of $W$ as a
function of time, averaged over 100 realizations of the disorder, for
a cylinder of radius $r_0=30$, cones with opening angles
$\theta=15^\circ$, $45^\circ$ and $60^\circ$, and the plane. In all
cases, the power-law behavior of the roughness with time, $W\sim
t^\beta$, is clear-cut, with a value of $\beta$ which is very close to
$1/3$, as expected.

\begin{figure}
\epsfig{file=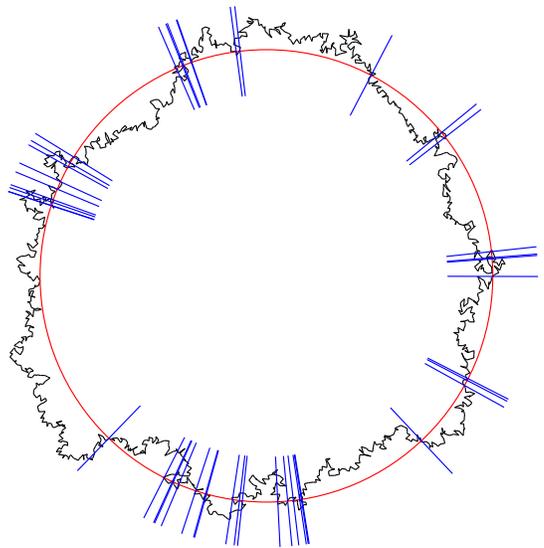,width=8cm}
\caption{Illustration of the procedure to estimate the surface
  correlation length $\xi(t)$. The profile is superimposed onto the
  best-fit circumference centered at the origin, and the intersection
  points are marked. The correlation length is estimated as the
  expected length of the patch to which a random point on the
  circumference belongs.}
\label{fig:illust_corr}
\end{figure}

The Family-Vicsek Ansatz also implies that the average roughness on
windows of size $\ell$ will scale as $w(\ell)\sim \ell^\alpha$ if
$\ell$ is smaller than the surface correlation length, $\xi(t)$.
Moreover, the three critical exponents are related via
$\alpha/\beta=z$.  In our case, direct measurements of the roughness
exponent $\alpha$ are involved, because distances along the curve
should be carefully computed.  In order to overcome this difficulty,
we have devised a novel technique to measure the correlation length,
which is illustrated in Fig.\ \ref{fig:illust_corr}. For a given
interface, we draw the best-fit circumference with centered at the
origin, and mark all the intersection points between the circumference
and the actual interface. They divide the circumference into a series
of $n$ patches or arcs, whose actual lengths
$\{\ell_1,\ell_2,\cdots,\ell_n\}$ on the cone are measured along the
azimuthal direction, being given by

\begin{equation}
\ell_i=\Delta\phi_i \left( r_0 + \left({\bar r}- r_0\right)\sin\theta \right),
\label{eq:ell}
\end{equation}
where $\bar r$ is the radius of the best-fit circumference.

\begin{figure}
\epsfig{file=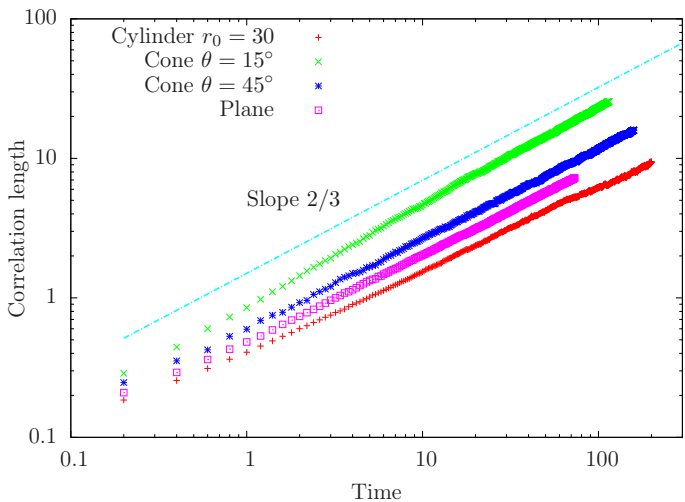,width=9cm}
\caption{Growth of the correlation length $\xi(t)$ for interfaces of
  different geometries: cylinder of radius $r_0=30$, cones with angles
  $\theta=15^\circ$ and $45^\circ$, and a plane. In all cases
  $\xi(t)\sim t^{1/z}$ with $1/z$ very close to $2/3$.}
\label{fig:corr}
\end{figure}

We can estimate the correlation length asking the following question:
if we choose a random point on the circumference, what is the expected
length of the patch on which it stands? On average, this value will be
given by

\begin{equation}
\xi \equiv {\sum_i \ell_i^2 \over \sum_i \ell_i} .
\label{eq:ell_average}
\end{equation}
Notice that this value does not correspond to the average value for
the patch lengths. The behavior of this correlation length $\xi$ is
shown in Fig. \ref{fig:corr}, where we can see that it follows a
power-law, with exponent close to the KPZ value $1/z=2/3$ in all
cases.  Thus, we have checked the first claim, that the interfaces on
cylinder, cones and plane, in all cases show the critical exponents of
the KPZ universality class.

\subsection{Radial Fluctuations}

The KPZ universality class does not only entail the values of the
critical exponents. As discussed above, the radial fluctuations are
expected to follow one of the well known Tracy-Widom probability
distributions. In the case of a ball on a random metric over the
plane, it was shown in Ref.\ \cite{Santalla_NJP15} that they indeed
follow the Tracy-Widom statistics for the Gaussian unitary ensemble
(TW-GUE).

We have developed an extension of the analysis in
\cite{Santalla_NJP15} in order to obtain the radial fluctuations
histogram for interfaces following the Huygens equation
\eqref{eq:huygens} on random conformal deformations of a base
Riemannian manifold, Eq.\ \eqref{eq:random_metric}, assuming that a
growing circumference is a solution of the aforementioned
Eq.\ \eqref{eq:huygens}. Along the simulation procedure described at
the beginning of this section, the radial data are stored along with
their time tag. We consider all pairs $(t_i,r_i)$, from different
noise realizations and times, and fit them to a linear form
$r_i=\varrho+vt_i$, whereby constant values $\varrho$ and $v$ are
obtained. Then, we fit the fluctuations to a power law with time,
namely,

\begin{equation}
\left( r_i-(\varrho+vt_i) \right)^2 = \Gamma^2 t^{2\beta}.
\label{eq:radial_fluct}
\end{equation}
Using the ensuing values of $\Gamma$ and $\beta$ (=1/3), we finally
extract the rescaled radial fluctuation $\chi$ as

\begin{equation}
\chi_i \equiv {r_i - (\varrho+vt_i) \over \Gamma t_i^\beta}.
\label{eq:chi}
\end{equation}
Notice that this $\chi$ variable is invariant under affine changes in
the radii $r$. We then obtain the histogram for these $\{\chi_i\}$ and
further normalize it, in order to have a distribution with zero mean
and unit variance. The theoretical prediction is that these histograms
will correspond to the TW-GOE and TW-GUE distributions in the extremes
of our family of surfaces: TW-GOE for the cylinder ($\theta=0$) and
TW-GUE for the plane ($\theta=\pi/2$).

These measurements have been carried out in three cases: (A) a
cylinder with $r_0=15$, for which we run 500 noise realizations and
gather all the points obtained from 1000 snapshots in the time
interval $t\in [1,10]$ for each noise configuration, giving a total of
$7\cdot 10^7$ points; (B) a cone with $\theta=15^\circ$, 500
realizations and 500 snapshots for each one with $t\in [100,200]$, a
total of $3\cdot 10^7$ points; (C) a cone with $\theta=45^\circ$, 100
realizations and 1000 snapshots for each one with $t\in [10,80]$, a
total of $4\cdot 10^8$ points.

\begin{figure}
\epsfig{file=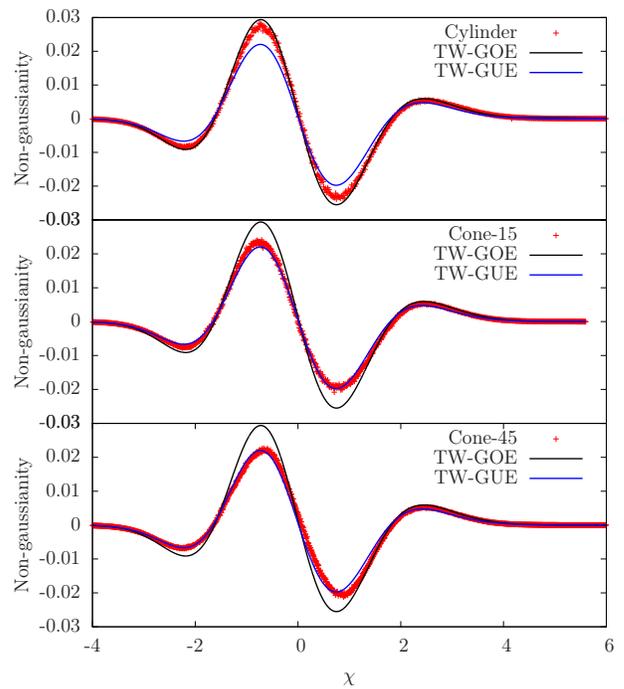,width=8cm}
\caption{Difference with a Gaussian (non-Gaussianity) of the radial
  fluctuations of interfaces grown with Huygens
  Eq.\ \eqref{eq:huygens} on random conformal deformations of our base
  manifolds: top, cylinder with $r_0=15$; center: cone with
  $\theta=15^\circ$; bottom: cone with $\theta=45^\circ$. Each panel
  includes the non-Gaussianity of the TW-GOE and TW-GUE distributions,
  for easy comparison. Notice that the cylinder corresponds to TW-GOE
  statistics, as expected, while the cones follow TW-GUE
  statistics. Further numerical checks are discussed in the text.}
\label{fig:tw}
\end{figure}

Before giving a quantitative assessment, let us consider the
visualization of these results as shown in Fig.\ \ref{fig:tw}. Since
the TW-GUE and TW-GOE distributions are very close visually to the
normal distribution, we plot the difference with the normalized
Gaussian probability density function, $\rho(\chi)=(2\pi)^{-1/2}
\exp(-\chi^2/2)$, which we call here {\em non-Gaussianity}. The top
panel shows the non-Gaussianity as a function of $\chi$ for the exact
TW-GOE and TW-GUE distributions, and for the obtained radial
fluctuations on the cylinder with $r_0=15$, which fit closely the
TW-GOE distribution, as expected. The central and bottom panels show
the analogous data for the cone with $\theta=15^\circ$ (central panel)
and $\theta=45^\circ$ (bottom panel). In these two cases, the
empirical distribution fits closely the TW-GUE distribution, as we
know to be the case for the plane \cite{Santalla_NJP15}. But, of
course, this check is merely visual, and should be supplemented with
further numerical comparisons.

A more strict test is provided by the estimation of the third and
fourth cumulants of the distributions, normalized as the skewness and
the kurtosis, as shown in Table \ref{tab:cumulants}. The data for the
cylinder can be seeen to correspond approximately to the TW-GOE
distribution, while they fit the TW-GUE distribution for the cones in
all cases.

\begin{table}[h]
 \begin{tabular}{@{}ccc}
      & Skewness & Kurtosis \\
      \hline
      TW-GOE & 0.2934 & 0.1652 \\
      TW-GUE & 0.2241 & 0.0934 \\
      \hline
      Cylinder $r_0=15$        & 0.30 & 0.18 \\
      Cone ($\theta=15^\circ$) & 0.24 & 0.10 \\
      Cone ($\theta=45^\circ$) & 0.23 & 0.13 \\
      \hline
    \end{tabular}
  \caption{\label{tab:cumulants} Skewness and kurtosis of the radial
    scaled variable $\chi$, Eq.\ \eqref{eq:chi}, for different base
    manifolds, as compared to the exact TW values.}
\end{table}

Another interesting measure is provided by the {\em Kullback-Leibler}
(KL) divergence between the empirical histograms and the theoretical
distributions. The KL divergence $D(P||Q)$ between two probability
distributions $P$ and $Q$ is defined as the loss of information when
data samples from $P$ are assumed to stem from $Q$ \cite{Desurvire},
and can be regarded as a natural distance in the space of
distributions. It can be computed as

\begin{equation}
D(P||Q) = \int \mu_P \log\left({P \over Q}\right),
\label{eq:kullback}
\end{equation}
where $\mu_P$ is the measure induced by distribution $P$. Table
\ref{tab:kullback} shows the KL divergences between the empirical
$\chi$ distributions and the TW-GUE and TW-GOE distributions. It can
be seen that, on the cylinder, the radial fluctuations are more likely
TW-GOE, but on all cones the radial fluctuations are closer to TW-GUE.

\begin{table}[h]
 \begin{tabular}{@{}ccc}
       KL-Distance to: & TW-GOE & TW-GUE \\
      \hline
      Cylinder $r_0=15$    &  $2.7\cdot 10^{-5}$ & $2.5\cdot 10^{-4}$ \\
      Cone ($\theta=15^\circ$) & $2.9\cdot 10^{-4}$ &  $8.3\cdot 10^{-5}$\\
      Cone ($\theta=45^\circ$) & $5.2\cdot 10^{-4}$ &  $2.6\cdot 10^{-4}$ \\
      \hline
    \end{tabular}
  \caption{\label{tab:kullback} Kullback-Leibler (KL) divergences,
    Eq.\ \eqref{eq:kullback}, between the empirical $\chi$
    distributions and the theoretical TW-GOE and TW-GUE
    distributions.}
\end{table}

\FloatBarrier

\section{Growth, geometry, and topology}
\label{sec:analysis}

The numerical simulations discussed in the previous section allow us
to extract several hypothesis. First, Huygens propagation on random
conformal deformations of cones of different opening angles are shown
to fall into the KPZ universality class, for all opening angles. We
can also conjecture that, on the cylinder, the radial fluctuations
follow TW-GOE statistics, while for all the cones with $\theta>0$ we
obtain TW-GUE. This conjecture fits well with the results of
\cite{Carrasco_NJP.14,Halpin-HealyJSP15}, where it was shown that
growth in a band geometry whose substrate expands at a constant rate
in time follows the TW-GUE distribution. In our geometric setting, an
expanding substrate is similar to a cylinder with a growing radius,
i.e., a cone.

These results require some theoretical explanation, which we will
attempt within our Riemannian geometry framework. Let us recall that
Huygens equation \eqref{eq:huygens} is {\em covariant}: solutions
obtained using one coordinate chart can be mapped into solutions
obtained using a different coordinate chart. The base metric tensor
for all our surfaces, in a polar chart $(r,\phi)$, has the form

\begin{equation}
g(r,\phi)=\begin{pmatrix}
1 & 0    \\
0 & f(r) \\
\end{pmatrix},
\end{equation}
where $f(r)=(r_0+(r-r_0)\sin\theta)^2$. If $\theta\neq 0$, an affine
change of coordinates,

\begin{align}
r \to\qquad &\hat r =r -r_0+ \frac{r_0}{\sin\theta},\\
\phi \to\qquad & \hat\phi=\frac{\phi}{\sin\theta},
\label{eq:reparam}
\end{align}
renders the metric Euclidean. Notice that this corresponds to viewing
the cone as a plane from which a wedge of angle $2\pi(1-\sin\theta)$
has been removed. For $\theta=0$, of course, the change of variables
\eqref{eq:reparam} becomes singular. And, as noted in the previous
section, an affine transformation in the $r$ coordinate will {\em not
  change the $\chi$ distribution}.

Let us now turn our attention to the conformal noise imposed upon the
base metric. Since we assume it to have vanishing short-length
correlations, we can safely assume that it will be invariant under
coordinate changes of any kind. Combining both statements we find
that, if $\theta\neq 0$, {\em the radial fluctuations for growth of
  any cone must have the same form as in the Euclidean case}. The same
argument can not be applied to growth on the cylinder, since in that
case the metric factor $f_{cyl}(r)=r_0$, and no affine change of
coordinates in $r$ will map it to the Euclidean case $f_{Euc}(r)=r$.
Notwithstanding, please notice that our argument does not entail that
the cylinder and the plane must have different fluctuations.

The difference between the cylinder and the rest of the cones is,
moreover, topological. All cones are homeomorphic to the plane, while
the cylinder is not. In fact, on the cylinder the Huygens equation is
applied in a different way. For any cone, we can start with an
infinitesimal circumference around the vertex and produce balls around
it. In the cylinder, we must start with a curve which is not
homotopically equivalent to a point, because it will wrap around the
manifold. But this difference by itself does not allow us to assert
that growth on the cylinder will possess different kinds of
fluctuations, since the cylinder can be smoothly {\em completed} with
a lower lid, thus rendering our initial circumference homotopically
trivial. Thus, the difference between TW-GUE and TW-GOE behavior does
not stem from the homotopy class of the initial curve.


\section{Conclusions and Outlook}
\label{sec:conclusions}

We have investigated the universality subclass structure of the KPZ
class in a Riemannian geometry setting for disordered substrates.  We
have studied the statistical properties of {\em Huygens} interfaces on
random metrics, see Eq.\ \eqref{eq:huygens}. A Huygens interface is
defined as a the propagation of an initial simple closed curve on a
certain manifold, always following the local normal direction with
unit speed. The metrics studied were conformal random deformations of
a certain set of base manifolds: the Euclidean plane, cones of
different opening angles, and a cylinder. In the planar case, it had
already been shown \cite{Santalla_NJP15} that the interfaces follow
KPZ statistics, with TW-GUE radial fluctuations. We have shown how KPZ
statistics are found in all other manifolds, with TW-GUE fluctuations
for the cones and TW-GOE for the cylinder. There is no intermediate
subclass between these two.

A theoretical explanation of this fact has been put forward, based on
the notion that the Huygens equation is {\em covariant}, i.e., it can
be studied in any possible coordinate chart. All cones with non-zero
opening angle are homeomorphic to the Euclidean plane, but not to the
cylinder. Moreover, we have written down the explicit non-singular
change of coordinates between the cones and the plane and shown that
it has no effect on the statistical properties of the radial
fluctuations of the interfaces, thus proving that all cones should
present TW-GUE statistics. This result fits very well with the results
of \cite{Carrasco_NJP.14,Halpin-HealyJSP15}, where it was shown that
KPZ systems in band geometry with an expanding substrate also feature
TW-GUE statistics.

Our work opens up many possibilities: what are the statistical
properties of the covariant KPZ equation on a generic manifold? Or,
alternatively, which are the statistics of the Huygens equation on
random deformations of a certain base manifold? In this case, we
expect a far richer set of possibilities. The topological argument
described in Sec.\ \ref{sec:analysis} suggests a possible methodology
in order to extract the radial fluctuations when the manifold is
homeomorphic to either the cylinder or the plane. But it leaves open
the question regarding the existence of new flavors or subclasses of
the celebrated KPZ universality class.


\begin{acknowledgments}
We acknowledge fruitful discussions with S.C.\ Ferreira and
K.A.\ Takeuchi. This work was funded by MINECO (Spain) Grants
Nos.\ FIS2012-33642, FIS2012-38866-C05-01, and FIS2015-66020-C2-1-P.
A.C. acknowledges financial support from the EU grants EQuaM
(FP7/2007-2013 Grant No. 323714), OSYRIS (ERC-2013-AdG Grant
No. 339106), SIQS (FP7-ICT-2011-9 No. 600645), QUIC
(H2020-FETPROACT-2014 No. 641122), Spanish MINECO grants (Severo Ochoa
SEV-2015-0522 and FOQUS FIS2013-46768-P), Generalitat de Catalunya
(2014 SGR 874), and Fundaci\'o Cellex.
\end{acknowledgments}


\end{document}